\documentclass[12pt]{article}
\usepackage[cp1251]{inputenc}
\usepackage{indentfirst}
\usepackage{graphicx}
\usepackage{subfigure}
\usepackage{amssymb}

\begin{document}



\title{Depths of the relief compensation  and the anomalous structure of crust and mantle of Mars}

\author{N.A.Chuikova\thanks{chujkova@sai.msu.ru}, L.P. Nasonova
\thanks{nason@sai.msu.ru}, and T. G. Maximova \\
Lomonosov Moscow State University,  \\ Sternberg  Astronomical Institute, \\
{\small Moscow, Universitetskii prospect, 13, 119991, Russia}}
\date{}

\maketitle
\medskip



\medskip

\medskip

PACS: 96.30. Gc  96.12.-a   96

\begin{abstract}
   The contribution to the   gravity  from the  Mars’ relief (topography) and the density jump at the Mohorovicic
discontinuity (${\bf M}$) in the quadratic approximation have been derived. The problem of determination  of possible
depths
of compensation for relief’s harmonics of different  degree and  order  have been solved. It is shown, that  almost all
 compensation of a relief is carried out in a range of depths of  0-1400 km. For various  relief’s
inhomogeneities  the  compensation is most probable  at the depths corresponding to the upper crust
 ($\bar d =4.5 \pm 3.7$ km),
to crust-mantle transition layer ( $\bar d =78\pm 24$ km); to lithospheric boundary ($\bar d =200\pm 34$ km); to
upper-mean mantle transition
layer ($\bar d =400±70$ km); to mean-lower mantle transition layer ($\bar d =1120\pm  180$ km). The lateral distribution of
compensation
masses is determined of this depths, and maps are constructed. The possible stresses in crust and  mantle of Mars
are evaluated. They reach $10^8$ Pa. It is shown that  relief’ anomalies of volcanic plateau Tharsis and symmetric
formation in east hemisphere  could arise and be supported dynamically by two plumes of melted mantle substance,
enriched by fluids.  The plumes have their origins on the boundary depths  of lower mantle.
\end{abstract}

\bigskip

{\it Keywords:}    Mars, gravity, isostatic compensation of relief, internal structure, crust, mantle, plumes, planetary dynamics.

\newpage

\section{Introduction}

\medskip
\qquad One of riddles of modern planetology is the  solution of the question, what forces support significant  global
 variations  of a relief heights of terrestrial planets which are  under the  external  gravity field and atmospheric
effects should be aligned  by the   process of denudation. One of the most interesting planets for research of this
question is Mars, a range of variations of  the Martian surface relief,  according to all available information Mars
Orbiter Laser Altimeter (MOLA)\cite{Zuber}, reaches 44 km, i.e. an order of size 0.013 in comparison with mean planetary
radius $R$ = 3389.5 km. For the Earth where the maximum range of variations of the relief heights led to uniform
density, is an order 0.002 $R$,  it has been shown  by us  \cite{Ch_2006}, that considerable anomalies of an internal gravitaty
field and a field of crust‘s stresses  can support existing differences of heights, despite of denudation’s processes.
 Such considerable gravity anomalous  in a crust  arise that basically the depth distribution of anomalous masses
has a dipolar pattern  owing to process of isostatic compensation. So, for example, opposite (in sign) Mohorovicic
({\bf M}) surface heights anomalies correspond to the  relief heights. Especially such conclusion can be fair for Mars,
 where a range of  relief heights variations  10 times more. Therefore, the investigation of the  global density
structures of Mars and comparison with  Earth are of   great  scientific interest.

\qquad Recently there were many observation data as on research of Martian topography \cite{Zuber}, and the  new models of a  Mars gravity
field,  derived from 5 years of monitoring by  Mars Global Surveyor  (MGS) \cite{Yuan,Konopliv}. These results have
allowed to draw some conclusions about the structure of a Martian crust \cite{Neumann}. So, in \cite{Neumann} the model of surface ${\bf M}$
 for Mars
(i.e., possible the crust-mantle transition layer ) on the basis  of Mars’ gravity  after  taking into  account  the
contribution of  the topography has been constructed. Similarly  the ${\bf M}$ model from earlier  data has been constructed
in \cite{Zarkov_91}.
This  problem was solved  in the linear approximation. However, our similar studies of density structure of the
Earth’  crust  showed that the linear approximation is insufficient for the accurate accounting of the contribution
of the crust  boundaries in the internal and external gravitaty fields, it is necessary to take into account the
quadratic  members \cite{Ch_2006,Nas_2007,Ch_2007}. Thus, the account of quadratic members from  the  relief’ expansion of  the degree N brings the additional contribution to harmonics of potential of degree n=0-2N,  with this   contribution increasing  with the growth of  n. Especially the account the quadratic  terms  is essential in dipole  (by  depth) distribution of anomalous masses. So, if the linear contribution to the external gravitational field is mainly correlated  isostatically with relief heights,  compensated on M, then the contribution from the account  quadratic terms  correlates with the squares of heights, i.e., is positive everywhere, both for continents, and for oceans, with  its  order of magnitude being comparable with the linear contribution. It turned out, that for some regions of the Earth the total contribution has  an opposite   sign as compared to the   linear contribution, which can considerably distort character of the  interpretation of satellite data.

\qquad In this paper, we consider  two methods of obtaining the   expansion coefficients of surface ${\bf M}$ for Mars, estimate
the contribution of the  relief  and the density jump (contrast)  on ${\bf M}$ to the  gravitational field of Mars  in the
quadratic approximation,   we compare   the   received results  among themselves and with the  appropriate
results for the Earth,  and  made estimates of the  possible distributions of depths of  compensation for the
relief mass   and  the density anomalies  and  that of stresses in the  crust and mantle  of Mars.

\bigskip
{\large \bf 2.\quad The calculation of the contribution from the Mars relief  and the  density jump at the  ${\bf M}$ discontinuity to the  gravitaty field in a quadratic approximation}
\medskip

\qquad In the linear approximation   the laterally  distributed anomalous masses are represented as a simple layer of
  continuous density distributed on a sphere. In this case, there is a linear relation between the coefficients of
the  expansion of a  simple layer density in terms of  spherical harmonics and the Stokes  constants caused  by the
the layer’s contribution \cite{Ch_2006}:

\begin{equation}\label{1}
\left\{ { \Delta C_{nm}^{(s)}}\atop{\Delta D_{nm}^{(s)}} \right\}=\frac{3}{2n+1} \left( \frac{R_s}{R_0}\right)^3 \frac{
\Delta\sigma_s}
{\bar \sigma } \left( \frac{R_s}{a}\right)^n
 \left\{{ a_{nm}^{(s)}}\atop { b_{nm}^{(s)}}
\right\},
\end{equation}

where $\Delta C_{nm}^{(s)}$, $\Delta D_{nm}^{(s)}$  are contributions to Stokes constants; $R_s$ is a mean radius of a layer s; $R_0 = 3389.5$ km,   $\bar \sigma = 3.93$ g/cm$^3$ and $a=3396$ km are mean radius,  density  and  mean radius at the equator  of the accepted  Mars reference ellipsoid, respectively \cite{Zuber,Yuan};

  $\Delta\sigma_s H_s(\varphi,\lambda)$ =
    $\Delta\sigma_s \cdot\Sigma_{n=1}^N \Sigma_{m=0}^n (\bar A_{nm}^{(s)} \cos m \lambda   +\bar B_{nm}^{(s)}\sin m\lambda)\bar P_{nm}(\sin \varphi)$ is a  representation of a simple layers‘ density as expansion series of  normalized  spherical harmonic  of degree  $n \leq N$; $ \Delta\sigma_{s} $, $H_s$    are mean density and  layer’s heights  relatively to the mean radius $ R_s $ and $ \left\{ {a_{nm}^{s}}\atop{b_{nm}^{s}}\right\} =\left\{ {a_{nm}}\atop{b_{nm}}\right\}_1=1/R_s \left\{ {\bar A_{nm}^{(s)}\atop{\bar B_{nm}^{(s)}}}\right\}$  are the normalized spherical harmonic coefficients of $ h_{s}=H_{s}(\varphi,\lambda)/R_s $.    The contribution of  layer’s  masses  to exterior gravitational potential is  then  defined  as follows:

$$
 \Delta V_e(r,\varphi,\lambda)=\frac{fM_0}{r}\! \sum\limits_{s} \sum\limits_{n=1}^N\!\left( \frac{a}{ r} \right)^n\! Y_n^{(s)}(\varphi,\lambda),
$$
$$
Y_n^{(s)}(\varphi,\lambda)=\sum\limits_{m=0}^n(\Delta C_{nm}^{(s)} \cos m\lambda+ D_{nm}^{(s)} \sin m\lambda) \bar P_{nm}(\sin\varphi).
\eqno (2)
$$
where $M_0$ is Martian mass.

\qquad In reality the relief masses and especially the anomalous masses, caused by density jump on ${\bf M}$, are not simple
spherical layers, but are distributed in height   relative to the reference
ellipsoid $r_e$. In this case  the coefficients of development (2) can be found by integration over relief masses:

$$ \left\{ { \Delta C_{nm}^{(s)}}\atop{\Delta D_{nm}^{(s)}} \right\}=\frac{\Delta\sigma_s}{(2n+1)M_0}\int\int\int r'^{(n+2)}\bar P_{nm}(\sin\varphi')\cos m\lambda'dr'd\lambda'd\sin \varphi' ,
$$
 where
$$ \int r'^{(n+2)}dr'=$$
$$\frac{r_e^{(n+3)}}{n+3}\left[ \left(1+ \frac{H_s}{r_e} \right)^{(n+3)}-1\  \right]
 \approx  R_s^{n+3}\left[ \frac{H_s}{R_s}+\frac{n+2}{2}\left( \frac{H_s}{R_s}\right)^2 +\alpha(n+2)\frac{H_s}{R_s}\bar P_{2}(\sin\varphi) \right]$$
in a quadratic approximation;  $r_e=R_s(1-\alpha \bar P_{2}(\sin\varphi)) $;  $ \alpha=\frac{2}{3}e $;
 and $ e=1/199.5 $ is  oblateness of the reference  ellipsoid,  for  which value    the  hydrostatic oblateness \cite{Zarkov_93} was taken.

\qquad Therefore, if one takes into account quadratic terms and ellipsoidal structure of reference surface
some additional terms arise in formula (1), namely
$$ \left\{ {a_{nm}^{s}}\atop{b_{nm}^{s}}\right\} =\left\{ {a_{nm}}\atop{b_{nm}}\right\}_1 + \frac{n+2}{2}\left\{ {a_{nm}}\atop{b_{nm}}\right\}_2 + \alpha (n+2)\left\{ {a_{nm}}\atop{b_{nm}}\right\}_3, $$
  where the terms in braces $\{\}_1$, $\{\}_2$,  and $\{\}_3$
with the subscripts 1, 2, and 3 correspond to the expansion coefficients of the functions $ h_s $, $ (h_s)^2 $,  and  $ h_s \bar P_{2}(\sin\varphi) $
  respectively.  The formulas expressing the coefficients $ \{a_{n m}, b_{n m}     \}_2 $  and $ \{a_{n m}, b_{n m}     \}_3 $,    in
  terms of the linear terms $ \{a_{n m}, b_{n m}     \}_1 $,  have been derived by mathematically simulating symbolic computations in computer algebra systems \cite{Nas_2007}.

\qquad Expansion of heights of surface ${\bf M}$ by the first mode ( ${\bf M1}$ model) was obtained by the method of
successive
approximation procedure. At first step, harmonic coefficients of the  Martian potential  after taking into
account   the contribution  from the topographic masses  into the  potential  in the quadratic approximation were found,  and
the expansion coefficients of the heights  of ${\bf M}$ in the linear approach were determined on the basis of the
obtained coefficients. Having found   the contribution to potential of the quadratic  members of heights of ${\bf M}$,  the new
expansion coefficients of  heights of ${\bf M}$ in the linear approach again were determined. The process of calculation was
repeated  to the complete convergence of  results .

\qquad The expansion coefficients  of the surface ${\bf M}$ heights by the second method ( ${\bf M2}$ model) were obtained using the hypothesis of isostatic compensation toopographic heights at ${\bf M}$. Here  for all harmonics, we used the   same transfer multiplier: $ k=\Delta d/\Delta h =(\sigma_r/\Delta\sigma_M)(R_0/R_M)^3 $, where $ \sigma_r=2.8$ g/cm$^3$ - is the mean  density of the relief’ masses, $ \Delta\sigma_M = 0.6$ g/cm$^3$ - mean density  contrast on ${\bf M}$, $ \Delta d $ - are  the depths of the surface ${\bf M}$ relative to the level surface corresponding to an average depth $ \Delta \bar d = 45$ km ( $R_M = 3344.5$ km)  \cite{Neumann}, and $ \Delta h $ - the heights of the  Martian topographic relative  to the  hydrostatic ellipsoid, corresponding to $ R_0=3389.5$~km.

\bigskip
{\large \bf 3.\quad Determination   of the  depths of  the compensation of  the  topographic masses }
\medskip

\qquad
Since there have been no seismic surveys on Mars, the preliminary data on its internal structure were obtained on the basis of observational data on its  gravity and topography, as well as on the basis of some theoretical conclusions \cite{Neumann, Zarkov_91}. These conclusions are based on cosmogonic scenarios about the formation of terrestrial planets, the geophysical and geochemical information and  on high energy physics data.

\qquad One of the first  tasks  at studying of  Martian  internal structure  is  to determine  of a crust-mantle   interface and a possible  density  jump at  ${\bf M}$.  The authors of  \cite{Zarkov_91}, proceeding from a hypothesis about effectively thermal formation of   terrestrial planets,  come to the conclusion   about  a possible  Martian crustal thickness  of  150-200 km and a density  contrast  of $ 0.5$ g/cm$^3$. In the same paper, where, on the basis of Bouguer anomalies, the contribution  from  a relief to a gravity is  estimated   in the linear approximation, the model of depths ${\bf M}$  relative  an average  depth level  of 140 km is  produced.   In \cite{Bab} it is shown, that a crustal  thickness  can vary in the range of 50-150 km depending on the mineral composition and  temperature distribution.  In \cite{Neumann},  proceeding from   petrological   and geophysical  constraints,  concluded  that the  thickness of a crust is  not  more than 50 km. In the same paper the authors  present  the model of depths ${\bf M}$  relative  mean  depth of 45 km and  at  the density  jump $ 0.6$ g/cm$^3$.  In \cite{Konopliv}   on the basis of the degree correlation  between harmonic coefficients of  gravity and topography, the mean  level of  relief compensation depths of  $ 72\pm 18$ km is obtained.

\qquad In the observed   works, and also in similar examinations (for example, in \cite{Turc}, where the  average depth ${\bf M}$ calculated on the basis of the  Airy compensation model   for different regions, is equal $90 \pm 10$ km),  the  problem  of compensation of relief masses was considered  at one level only,   the level of  ${\bf M}$ discontinuity.  However,  analogous studies for the Earth \cite{Ch_2010} show, that in planet’ interiors  there can be several compensation levels, consistent   with the results obtained from  analysis of the Earth’s eigentones  and seismological data. Thus depths of compensation for  different  topography harmonics  proved to be   strongly  dependent on  the harmonic degree and  order. Therefore our first task was to determine the possible  depths of compensation   for  harmonics of different degrees and orders for expansion of heights of Mars’relief    relative to  the  hydrostatic ellipsoid.

\qquad The solution of this  problem should satisfy  a sistem  of  two equations,  where one of which reflects the consistency between  contribution  of  topographic and compensating masses and the observations,  and the other  equation reflects the equality of the pressures below the compensation  depth  to  the  hydrostatic pressure. The solution obtained for the  compensation depth  $ d_{nm} $  for an arbitrary  relief harmonic $ a_{nm} $ was defined as a result of the relationship:
$$
d_{nm}=R_0-R_M \left(  a_{nm}^{M1}/b_{nm}^{M2}   \right)^{1/n},
\eqno (3)
$$
where $ a_{nm}^{M1} $ and $ a_{nm}^{M2}  $  are the expansion coefficients of  heights of the surfaces  ${\bf M1}$ and  ${\bf M2}$ at the  fixed value of $ R_M $ (for  relief harmonic $ b_{nm} $   the  formula  is similar with  replacement  $ a_{nm} $  by  $ b_{nm} $).

\qquad As we see, solution (3) is possible, i.e. $ d_{nm}\geq 0 $ if $ 0\leq a_{nm}^{M1}/b_{nm}^{M2}\leq (R_{0}/R_M)^n$.

\qquad  If expansion coefficients of the relative altitudes  at  ${\bf M}$, received  by the first (${\bf M1}$) and second (${\bf M2}$)  method, do not satisfy  condition  (3) for any fixed  compensation  depth $ d_{nm}<R_0 $,  there should be a two-layer compensation. The  possible depths  of compensation  layers are determined  then from the analysis of the  results obtained for those  harmonics for which  there exists a solution (3). The final choice  is based  on the   principle of minimization  of deviation of the  internal structure of Mars from the hydrostatic equilibrium.

\bigskip
{\large \bf 4.\quad Results and disscussion }

\medskip
\begin{figure}[]
\subfigure[]{\includegraphics[width=0.22\textwidth,,
angle=-90]{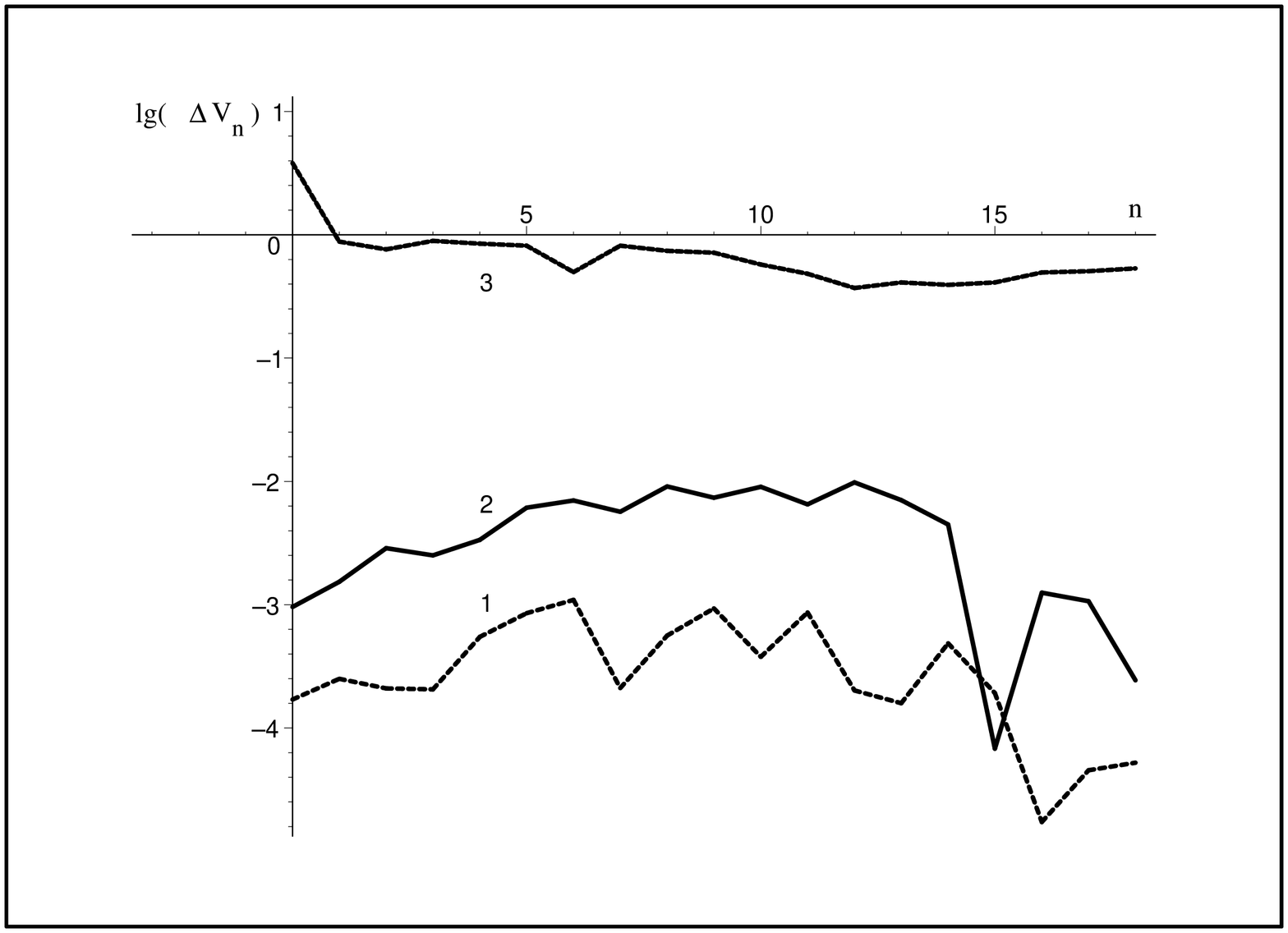}}
\subfigure[]{\includegraphics[width=0.22\textwidth,,
angle=-90]{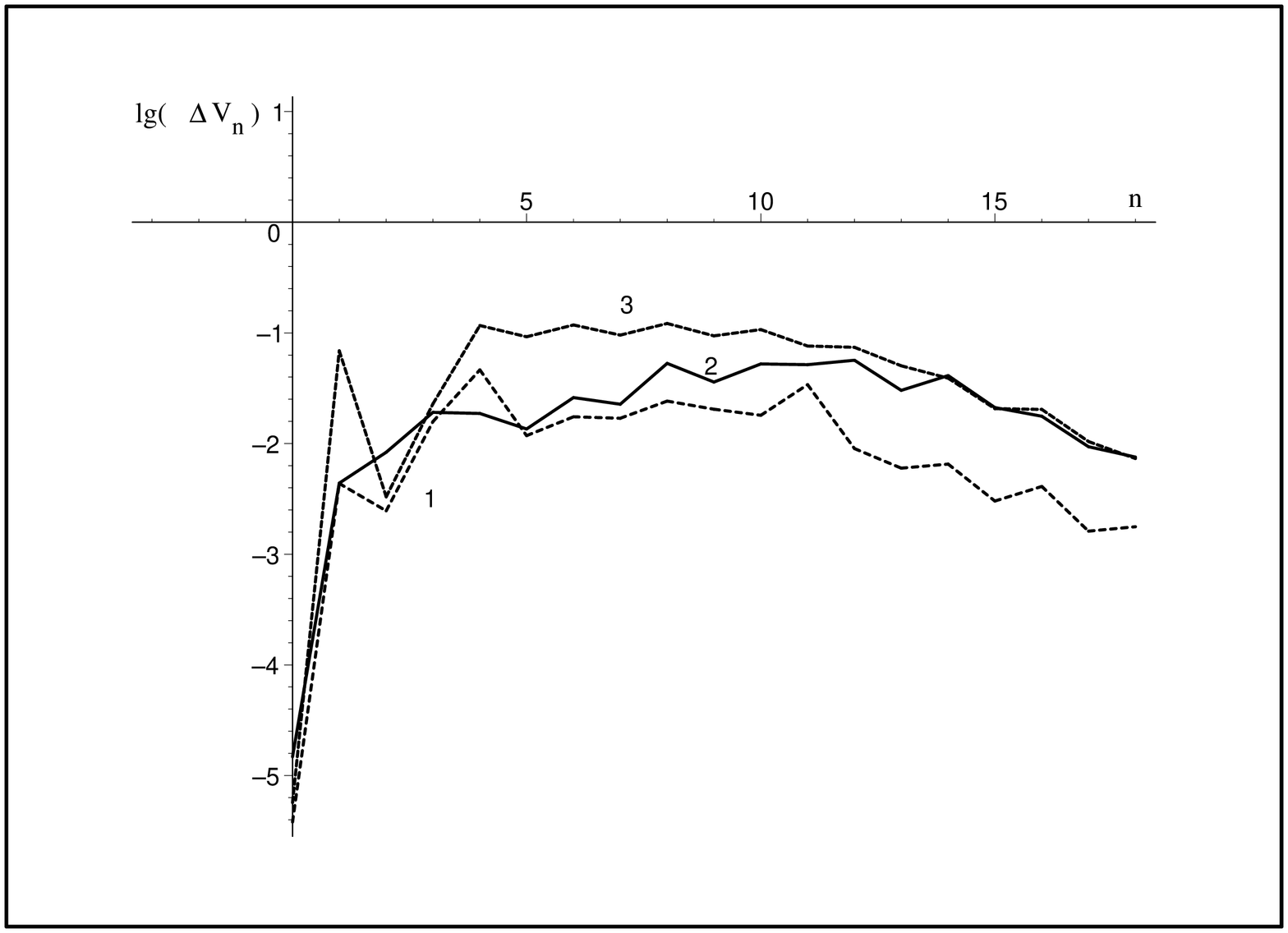}}
\subfigure[]{\includegraphics[width=0.22\textwidth,,
angle=-90]{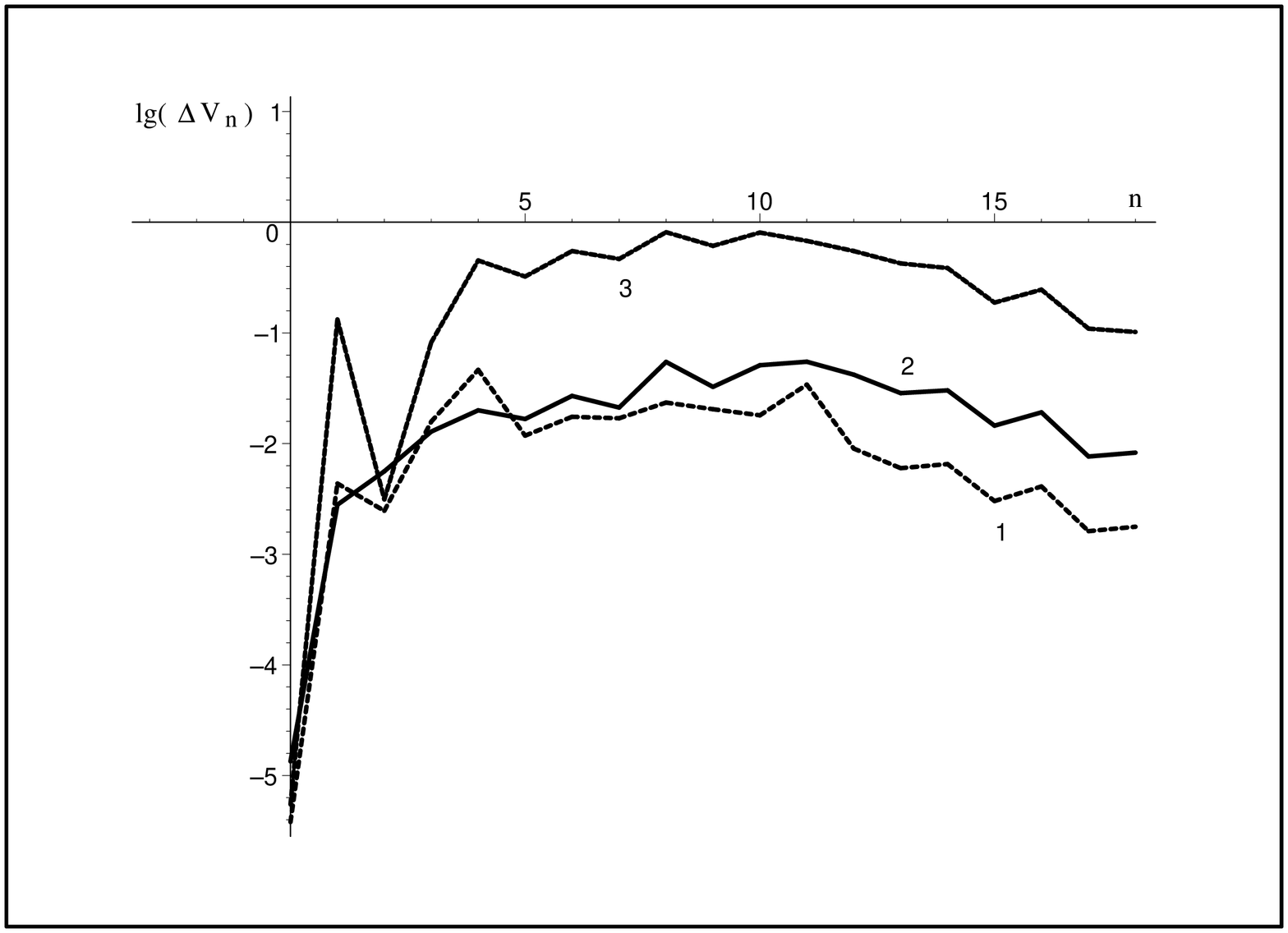}} \caption{\footnotesize Dependence of the
relative mean-square contribution of quadratic members to the
planet’s external gravitational potential on the expansion degree
$n$ (a)--- for the Earth; (b)--- for Mars, the ${\bf M1}$ model, and
(c)--- for Mars, the $\bf M2$ model. Notations: 1--- for topographic
masses; 2 --- for he density jump at $\bf M$; and 3--- for the total
contribution. } \label{Fig1}
\end{figure}

\qquad (1) In this study, we used algorithms and formulae  that express the expansion coefficients  in the   spherical functions of the square of some function of $(x = \sin \varphi, \lambda)$, given initially in the form of a similar expansion into spherical harmonics, through the coefficients of this initial  expansion \cite{Nas_2007}.
 Algorithm developed by us allows to receive these formulae for arbitrary  degree  $ N_{max} $ of initial  expansion. The final formulae for numerical calculations are obtained for  $ N_{max}=9 $
  inclusive, and they allow one to estimate  the contribution of quadratic members into the Stokes constans of the degree $ n=0...18 $.  According to \cite{Zarkov_91}, the decomposition to $N = 18$ is
   quite sufficient  to  identify  the main singularities of  Martian  topography   and  gravity. We used the contribution of the quadratic terms to the Earth’s and Mars’s  gravitational field from the relief masses and the crust-mantle density contrast at ${\bf M}$ to illustrate the results obtained.
 (Fig1) shows  the dependence on $n$  mean-square  contributions
 $ \Delta V_{n}=\sqrt{D_{n,2}/D_{n,1}} $
for   the relief   masses and density  jump  at  ${\bf M}$,  both separately and together. Here
 indexes 1 and 2     correspond to the linear  and    quadratic members, respectively; $ D_{n}=\sum \limits_{m=0}^{n} \left( (a_{nm})^2 + (b_{nm})^2 \right) $, $a_{nm}  $ and  $b_{nm}  $
are the harmonic coefficients of  presented expansion.

\qquad A comparison of Figs.1a, 1b, and 1c shows that, for Mars,  the contribution of the quadratic terms in the external potential  both of  topographic  masses, and  the density jump at ${\bf M}$,  on average, 10 times  order of magnitude larger than  for the Earth, and is comparable  to the linear contribution  on the order of smallness for 40\% of the coefficients. This is especially typical  for the ${\bf M2}$ model, for which the quadratic contribution exceeds  the linear contribution  for 20\%  of the  coefficients ( beginning  from $n = 1$), while for the Earth,   the quadratic contribution exceeds   the linear  contribution  only for  a few coefficients from  the  range under consideration. The  distribution histogram  for the  relative quadratic  contribution to the  Stokes constants (the ratio of the quadratic terms to linear) reaches its  maximum  (22\% of coefficients) for the relative square contribution equal 0.4 (for the Earth, these figures  are  16\% and 0.01, respectively \cite{Nas_2007}).

\qquad A comparison   of the  two arrays of  the expansion harmonic  coefficients  of heights   boundary    ${\bf M}$ shows, that determination of altitudes of ${\bf M}$ by the first method appears to be incorrect.  For example,  in  \cite{Ch_96}  for  the Earth it is shown, that the  conventional  practice for   determination  the  ${\bf M}$ depths  on the basis of  Bouguer anomalies is not confirmed  for almost  all harmonics of the  degree $n> 2$ (there is  no  correlation between the Bouguer  anomalies and  the ${\bf M}$ depths  obtained from   seismic data). From this work, and also from \cite{Ch_2003}  follows, that for the Earth  the transmission multipliers   decrease with  the growth of  the expansion degree   $n$ approximately   according  to a  linear law. Obviously, for a  thicker Martian crust  (which is  three times  thicker  in comparison with the Earth’s crust) the decreasing  law must  be  more strongly expressed.  However, a  comparison of the expansion coefficients  of the ${\bf M}$ heights obtained by the first and second method shows that the  transmission  factors, on the average, do not decrease with $n$, and even for some harmonics, they  increase. Apparently,  the Mars gravity anomalies essentially depend on the inhomogeneous  density  structures as well as  Martian crust,  and the   deeper layers.

\begin{figure}[]
\subfigure[]{\includegraphics[width=0.22\textwidth,
angle=-90]{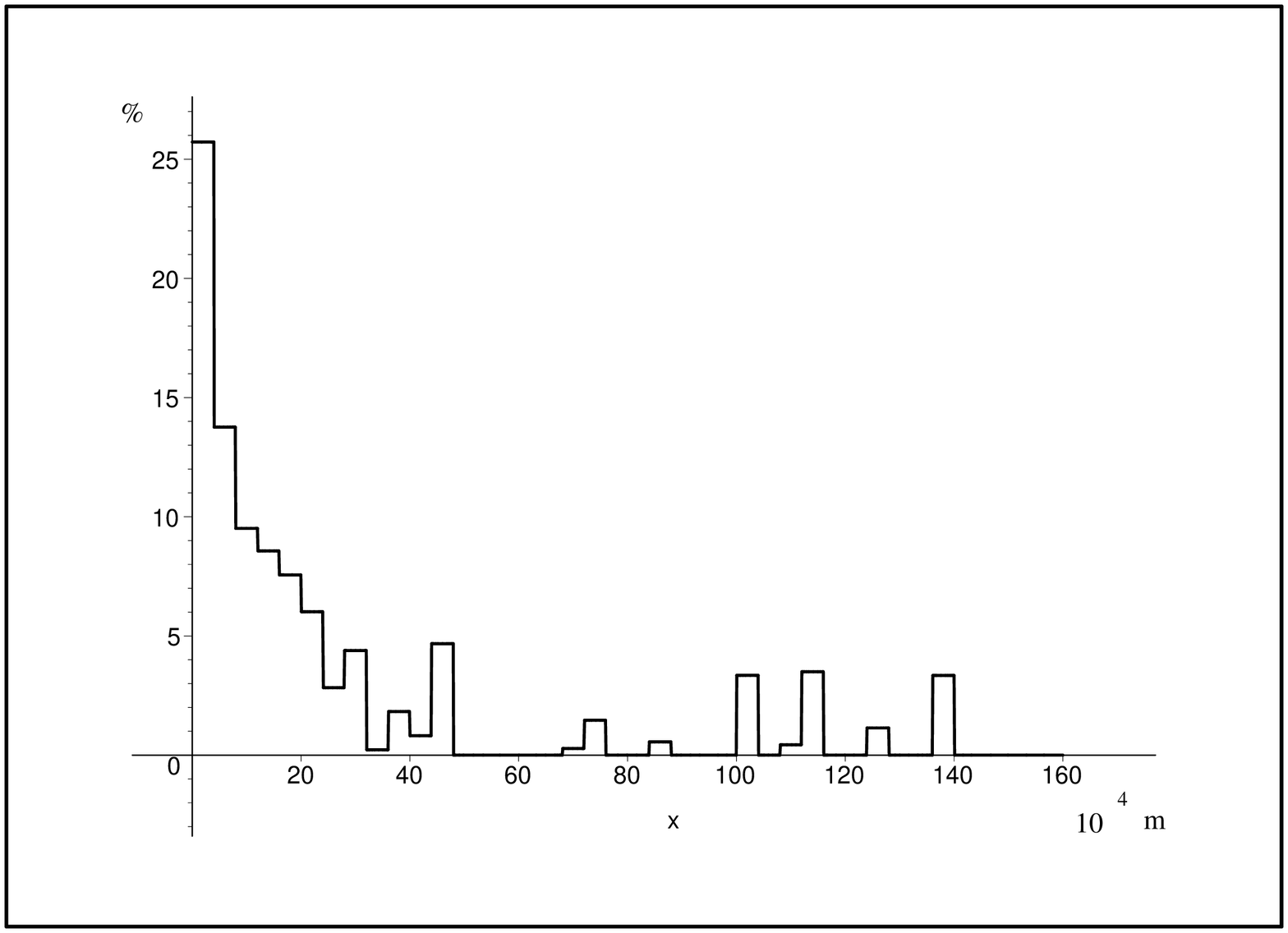}}
\subfigure[]{\includegraphics[width=0.22\textwidth,
angle=-90]{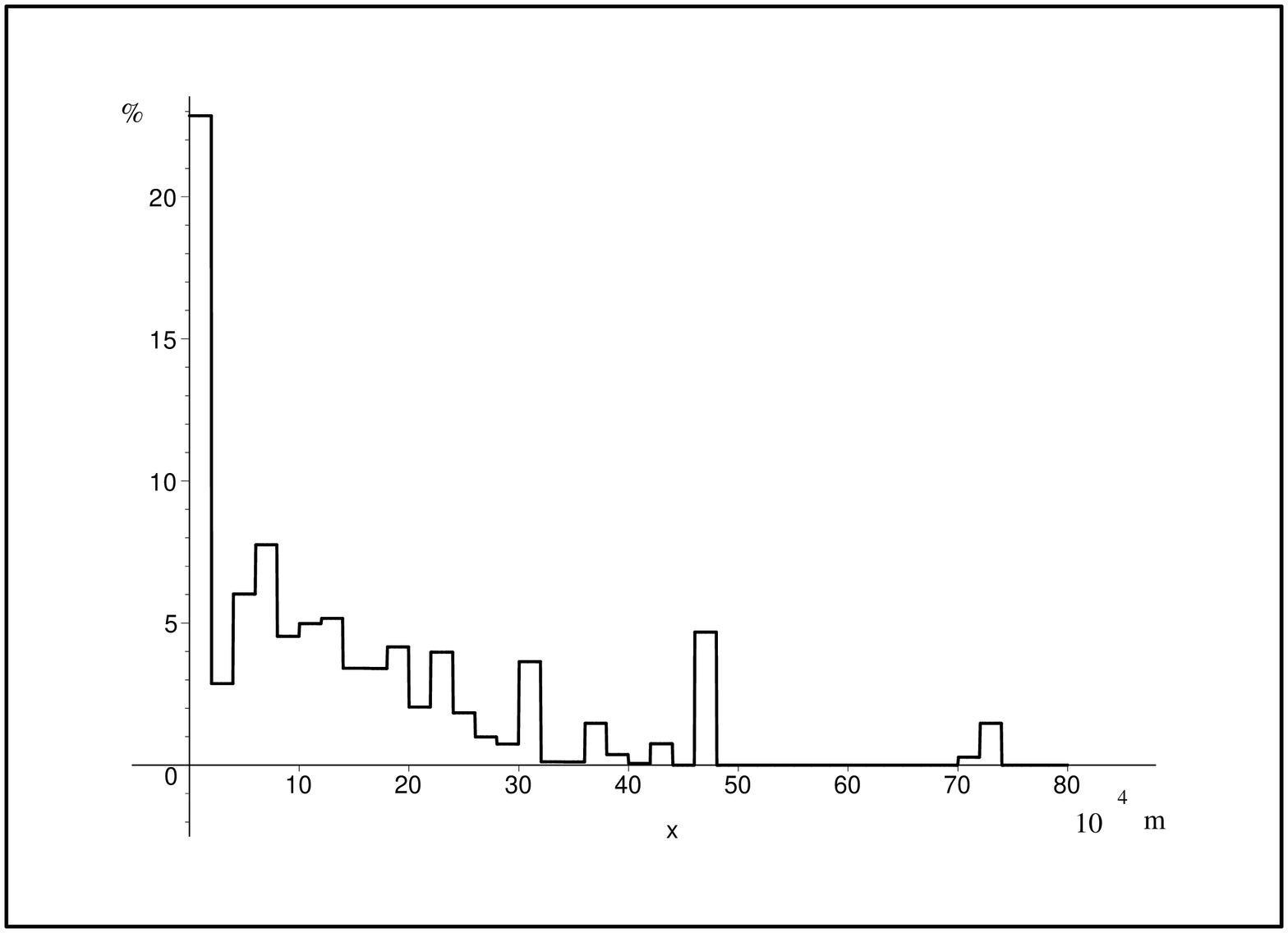}}
\subfigure[]{\includegraphics[width=0.22\textwidth,
angle=-90]{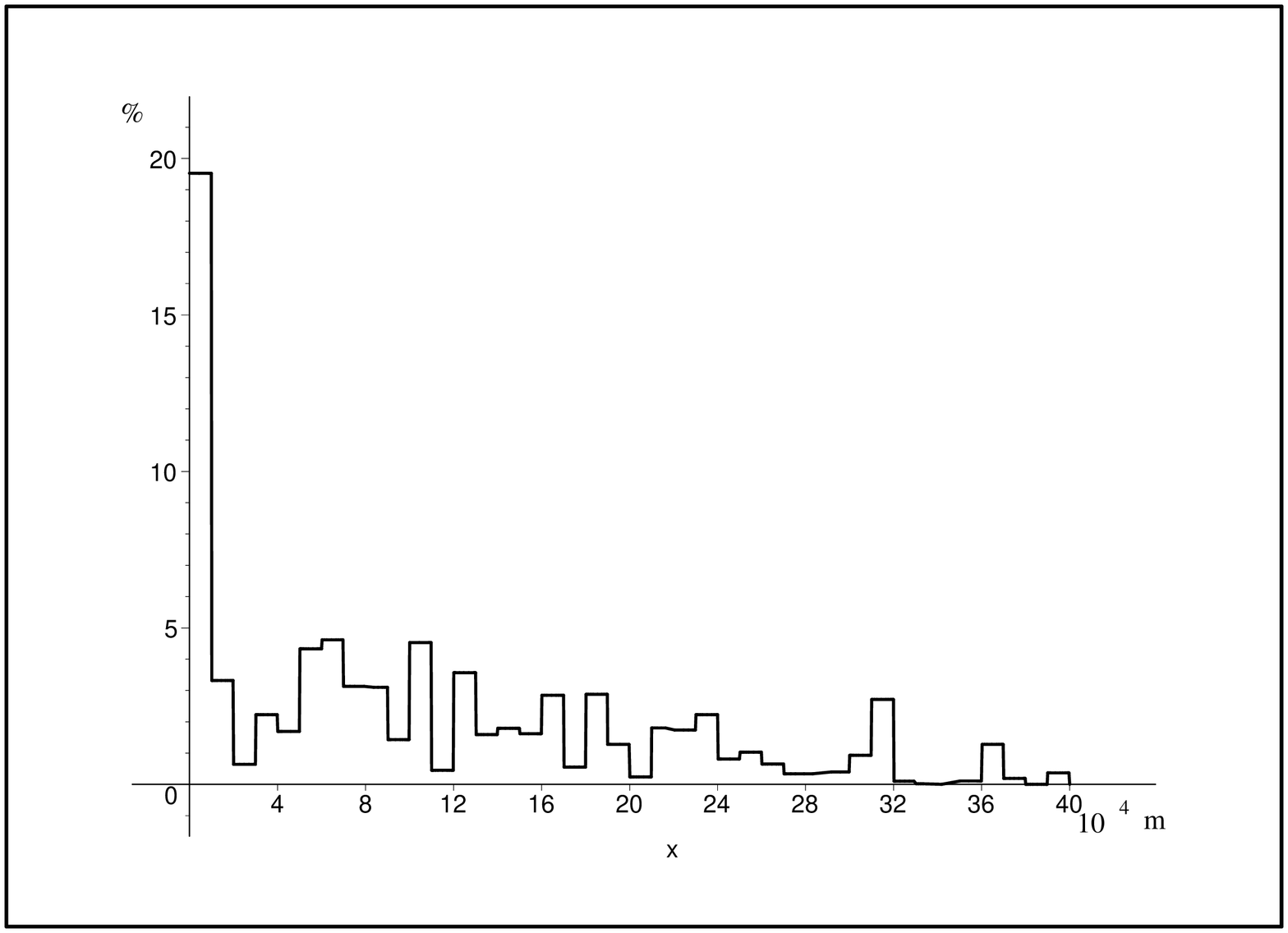}} \caption{ \footnotesize Histograms of the
distribution of the compensation depths for topography harmonics
with a step of (a) 40 km, (b) 20 km, and (c) 10 km. } \label{Fig2}
\end{figure}

\qquad    (2)  Figure 2 presents    distribution histograms  of the compensation  depths  of  topography harmonics    for the entire  depth range  $0 - 1500$ km
with  a step of  40 km (Fig. 2a), 20 km ( Fig. 2b)  and more detailed for the depth range  $0 - 390$ km  with a step of 10 km ( Fig. 2c),  obtained from   (3) for  $n\leq18$.  The histograms and mean  depths  were calculated taking into account weights corresponding to the amplitudes of considered topography  harmonics.
  An analysis of the  histograms results in the  following  conclusions:  almost all the topography is  compensated   in the depth  range   $0-1400$ km, and  about  20\% of compensation occurs in the upper crust (the   depth range $ d=0-20$ km, $\bar d =4.5 \pm 3.7$ km). Further it is possible to identify  several  main   layers of compensation: a boundary interface  crust-mantle ($d=50-130$ km, $\bar d =78 \pm 24$ km); the  lithospheric boundary ($d=160-240$ km, $\bar d =200\pm 34$ km);  upper- middle  mantle transition layer ($d=310-470$ km, $\bar d  =400±70$ km); and   middle-lower mantle  transition layer ($d=1000-1400$ km, $\bar d =1120±180$ km). Note that in each   layer the  compensation maximum  occurs usually near the  layer’s boundary. This may  speak about  variations in the heights of the boundaries separating layers of different densities.

\begin{figure}[]
\includegraphics[width=0.995
\textwidth ]{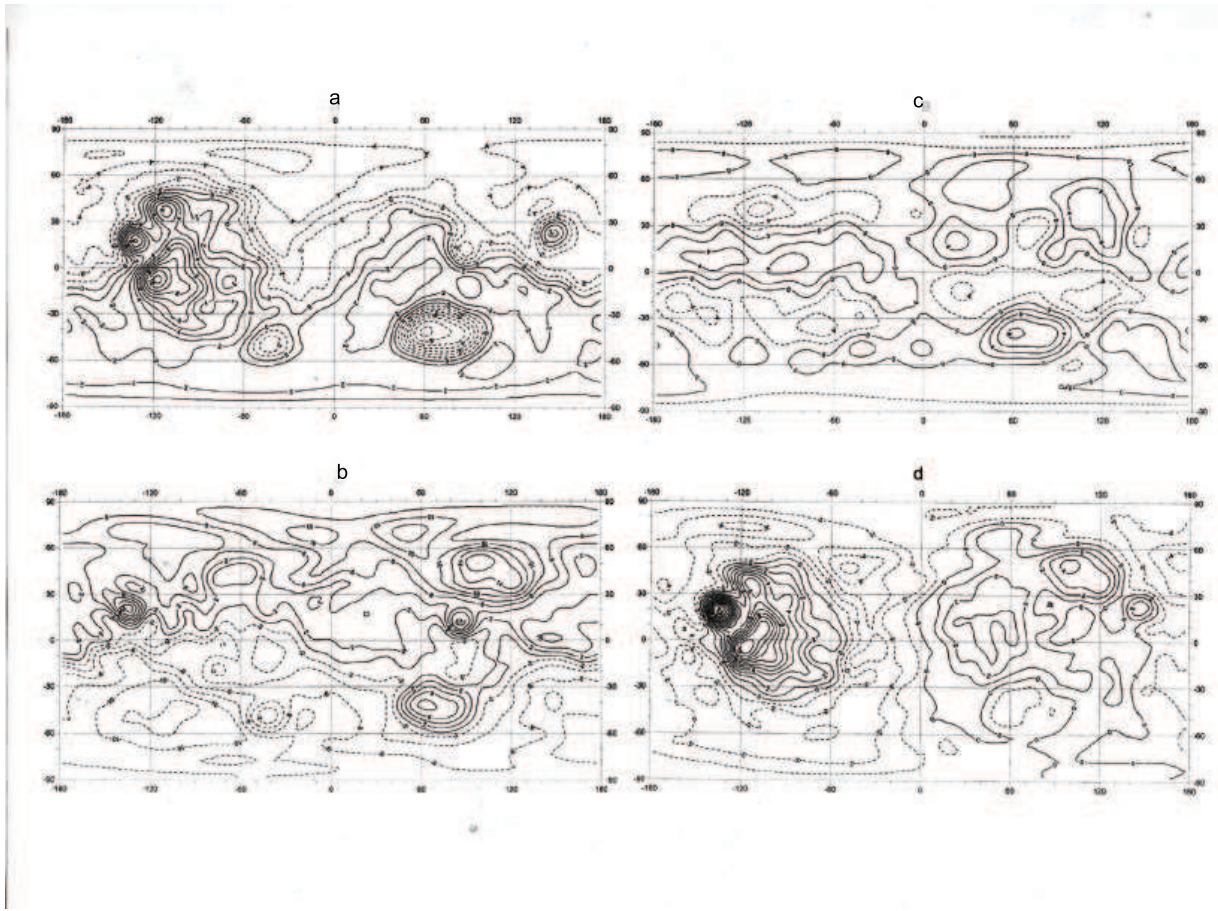} \hfill
\caption{ \footnotesize Anomalous structures of the crust of Mars
for the expansion to 18 degrees:
 (a) the height of relief relative to hydrostatic ellipsoid, the  isoline cross section is 1 km, the range of  variations  is ( -7.2, 11.5) km; (b) - density anomalies in the upper crust (at depths of 0 -- 20 km), the  isoline cross section is  2$\cdot10^6$, the range of variations  is  ( -13.0,17.5)$\cdot10^6$; (c) -- the density anomalies in the boundary layer of crust-mantle (at depths
of 50 -- 130 km), cross isolines is  2$\cdot10^6$, the range of
variations is ( -6.9, 8.4)$\cdot10^6$;   (d)  nonhydrostatic
pressure of the crust on the lithosphere (at  depths of 130 km), the
isoline cross section is $10^7$ Pa, the range of variations is (
-5.4, 14.5)$\cdot10^7$  Pa. }
 \label{Fig3} \end{figure}

\qquad As each   relief’ inhomogeneity  is characterized by a certain set  of harmonics, the maximum concentration of the  compensation for  this set in a certain range of  depths    can testify to the most probable depths of compensation of considered relief’ inhomogeneity.    Figure 3 and figure 4 present density maps  of the compensating masses, recalculated to a densities of a simple layer at mean depths 0, 78, 200, 400, 1120 km corresponding to the set of harmonics for layers with this mean  depths.
 For harmonics for which condition  (3) is not fulfilled,  two-layer compensation  at the selected depths are introduced. In this case for each harmonics and  all the  possible variants of compensation,   we defined  the weight function,  that is inversely proportional to the sum of absolute values of the  density variations, and by taking it into account,  we calculated  the density anomalies  at  considered depths. We have used  the presented procedure earlier in  \cite{Ch_2010} for determining the  density anomalies  in the crust and mantle of the Earth at the depths, selected on the basis of seismic data. The obtained  density distributions  are in good agreement   with the results, obtained on the basis of the spectral analysis of the  eigenfrequency normal modes for  the Earth. The application of this  same procedure for Mars allows one to count   on the reliability of the obtained results, although this is  only a  model  representation

\qquad A  comparison of the obtained   density distributions  allow   the following conclusions: (1) the dichotomy of the relief of Mars (Fig. 3a), caused by the first  degree  harmonics, in the main  is compensated  by   lava filling of the crust  of the northern hemisphere  plains (Fig 3b);  (2) the relief anomalies of the  Tharsis  volcanic plateau and the  symmetrical formation in the eastern hemisphere, caused mostly   by  second degree  harmonics, possibly,  may  have arisen and are dynamically supported  by  the presence of two plyumes  of the enriched by fluids molten   mantle substance,   with these plumes having their
origin  on the upper and lower mantle boundary (Fig. 4c) and decomposing into several branches  at  the tops of the  upper mantle (Figs. 4a,  4b). The penetration of these plumes through the mantle and the lithosphere became possible, apparently, after the impact of the large asteroid, which  have led to the topographic dichotomy  and  the formation of cracks in the crust and  mantle, through which the molten substance reached  the surface. The lift of the plumes’ light substance  was  later complicated  by the presence of downward  flows of the partially cooled heavy substance of lavas  in the process of gravitational differentiation (Figs  4a, 4b). Let us note  that the ascending current into the mantle, the stronger under the Tharsis, pushes aside around the descending masses, and in the symmetrical equatorial region the stronger downflow pushes aside around the light ascending masses (Fig.  4b). Figure  4b can also  be interpreted as the sagging of the  lithosphere under  Tharsis  and  its lifting around the plateau, and as the lift  of the lithosphere under the symmetrical equatorial region and lowering the surrounding areas, caused by the  load pressure of the  lithosphere’ masses.

\begin{figure}[b]
\includegraphics[width=0.99\textwidth ]{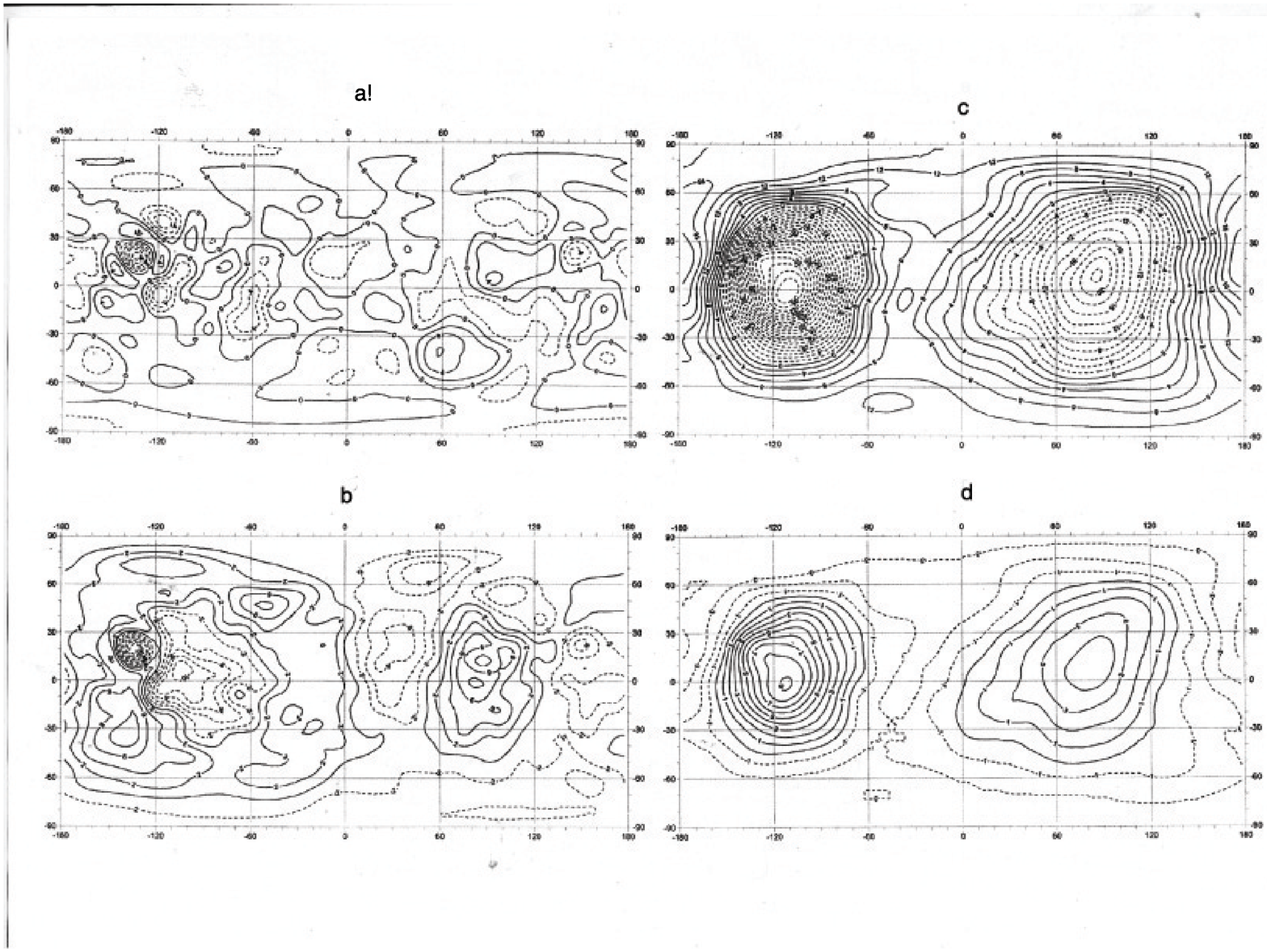}
\hfill
\caption{ \footnotesize Anomalous structures in Martian mantle for
an 18-degree expansion: (a) density anomalies in the lithosphere
boundary layer (at a depth of 160 -- 240 km); the contour interval
is 2$\cdot10^6$ kg/m$^2$; the variation range is (from –11.6 to
6.0)$\cdot  10^6 $kg/m$^2$; (b) density anomalies in the
upper-middle mantle transition layer (at a depth of 310 -- 470 km);
the contour interval is 2$\cdot10^6$ kg/m$^2$;
 the variation range is (from –16.3 to 9.5)$\cdot 10^6$ kg/m$^2$; (c) density anomalies in the middle-lower mantle transition layer
(at a depth of 1000 -- 1400 km); the contour interval is
2$\cdot10^6$ kg/m$^2$; the variation range is (from –39.8 to 20.0)$
\cdot 10^6$ kg/m$^2$; (d) nonhydrostatic pressure in the middle
mantle (at a depth of 600 km); the contour interval is $10^7$ Pa;
the variation range is (from –4.2 to 9.2)$\cdot 10^7$ Pa.}
\label{Fig4}
\end{figure}

\qquad Figure  4 shows that the volcanic craters of increased density (Olympus  and  craters of Tharsis plateau) have their sources in the upper-lower mantle interface layer at depths of $1120 ± 180$ km, Elysium  and smaller craters around  Hellas  Planitia,  in the upper mantle at depths of $400 ± 70$ km and at depths of $200 ± 34$ km, small craters between the Tharsis and Argyre, in the crust-mantle  transition  layer (Figure 3c) at depths of $78 ± 24$ km. The craters of impact origin (Hellas, Isidis, and  Argyre),  perhaps analogous to the lunar mascons of the increased density, extends  down to  depths of $200±34$ km and are surrounded by a ring  structures with reduced density (Figs 3c , 4a).
It is interesting to note  the  distributions of the density anomalies  that  are elongated  in longitude in the crust- mantle  boundary layer (Fig. 3c). Perhaps this is due to   the accumulation in this layer of lava and  fluid flows from the underlying mantle  layers, with different velocity distribution in longitude.  These flows can stretch along     the longitude also under the influence of Coriolis forces,  generated by the convective motions in the mantle near the  crust boundaries  to the north or south. The regions of negative anomalies, including subpolar  areas, may characterize the reserves of fluids, including water,  that   had  not time to reach the surface of Mars through the hardened crust.

\qquad It is possible to   explain  also Figure 3c  by the variations in the boundary  ${\bf M}$. There is no clear correlation of depths ${\bf M}$ with the structures of relief (with some exception for the Utopia,  Elysium, Hellas, Argyre and Alba because of the partial compensation on ${\bf M}$). Under the Great Northern plain, Utopia, Arabia Terra,  Terra Sirea, Terra Cimmeria and Prometheus Basin,  boundary ${\bf M}$ could be raised under the influence of  upward flows  from the mantle (Figs  4a,b), under  Alba Patera, in the equatorial eastern hemisphere and a southern part of Tharsis it was  omitted under the influence of the load pressure of crust.

\qquad Nonhydrostatic pressure in different layers of the crust and mantle,  produced by the pressure of  the overlying layers,  reach  their maximum  of   145 MPa in the lithosphere (Fig. 3d), and  decrease  and become smoother in the middle mantle (Fig. 4d). They must be balanced by the pressure of the underlayers  (because of   the  isostatic equilibrium condition) and characterize the distribution of the vertical and horizontal stresses (positive values  correspond to vertical compressive stresses and  horizontal tensile stresses, and vice versa).  A comparison of the maps shows that in the Martian mantle, there could    have been (or exist till  now), the convective motions, which, perhaps, was the source of nonhydrostatic stresses ( within  the dynamic approach). The absence of the  density and pressure anomalies below the depth of 1400 km   shows that the deeper Martian interiors are in a  well-established equilibrium, that also leads to the absence of convective motions in the core and, therefore, to the absence of the conditions for  a  hydromagnetic dynamo.

\medskip

\begin{center}

{ \bf \quad Conclusions}
\medskip
\end{center}

\qquad At an estimating the contribution of the  relief’masses  and  the  density  jump at ${\bf M}$  (Moho density contrast) to  Martian gravity, it is necessary to consider the  quadratic members. Eestimating the  ${\bf M}$ depths  from  Bouguer anomalies, which is consistent with the hypothesis  of the  homogeneous  structure of the Martian crust, contradicts the   data of the  analysis of transmitting coefficients   for Mars and the analodous   estimates for the Earth.  The Martian crust and mantle are  characterized by a nonhomogeneous distribution of density and stresses down to the depth 1400 km.  The topographic anomalies  of the Tharsis  volcanic plateau  and  the symmetrical formation in the eastern hemisphere, possibly,  have arisen, and  be dynamically supported  by   two plumes of  melted mantle substance, enriched by fluids. The plumes have their origins  at  the boundary  of the  lower mantle.


\end{document}